# Road Surface Friction Prediction Using Long Short-Term Memory Neural Network Based on Historical Data

Ziyuan Pu [1], Shuo Wang[2], Chenglong Liu[3], Zhiyong Cui[4], Yinhai Wang[5*]

[1] Smart Transportation Application and Research Laboratory, Department of Civil and Environmental Engineering, University of Washington, 101 More Hall, Seattle, WA, U.S.A.
[2] School of Computer Science and Engineering Nanjing University of Science and Technology 200 Xiaolingwei Street, Nanjing, Jiangsu, China.
[3] Key Laboratory of Road and Traffic Engineering of the Ministry of Education, Tongji University
4800 Cao'an Road, Shanghai, China
[4] Smart Transportation Application and Research Laboratory, Department of Civil and Environmental Engineering, University of Washington, 101 More Hall, Seattle, WA, U.S.A.
[5*] Department of Civil and Environmental Engineering, University of Washington, 121F More Hall, Seattle, WA, U.S.A.
[*] corresponding author: yinhai@uw.edu

**Abstract:** Road surface friction significantly impacts traffic safety and mobility. A precise road surface friction prediction model can help to alleviate the influence of inclement road condition on traffic safety, Level of Service, traffic mobility, fuel efficiency, and sustained economic productivity. Most related previous studies are laboratory-based methods which are difficult for practical implementation. Moreover, in other data-driven methods, the demonstrated time-series features of road surface condition have not been considered. This study employed a Long-Short Term Memory (LSTM) neural network to develop a data-driven road surface friction prediction model based on historical data. The proposed prediction model outperformed the other baseline models in terms of the lowest value of predictive performance measurements. The influence of the number of time-lags and the predicting time interval on predictive accuracy was analyzed. In addition, the influence of adding road surface water thickness, road surface temperature and air temperature on predictive accuracy also were investigated. The findings of this study can support the road maintenance strategy development and decision making, thus mitigating the impact of inclement road condition on traffic mobility and safety. Future work includes a modified LSTM-based prediction model development by accommodating flexible time interval between time-lags.

## 1. Introduction

Road surface condition has a great impact on road traffic mobility and safety [1–3]. Especially in the winter season, terrible road surface conditions could result in more traffic crashes and low level of service (LOS). The United States spends $2.3 billion annually to keep highways clear of snow and ice; in Canada, winter highway maintenance costs more than $1 billion [4]. Improving road surface condition monitoring systems and operations could result in fewer crashes, higher LOS, improved mobility, better fuel economy and sustained economic productivity [5]. As one of the direct measurements of road surface condition, road surface friction has a strong correlation with traffic accident risk [6]. Thus, in order to mitigate the impact of road surface condition on traffic safety and mobility, an efficient and cost-effective road surface friction prediction methodology is needed for these concerns.

Most previously prediction models for road condition related parameters predicting are developed based on laboratory tests. Shao et al. proved that the ice hazard only happened under both conditions based on field test data from seven countries [7]. They also tried to predict the ice condition based on air temperature, wind speed, and precipitation. However, the results showed great differences in different roadways. Samodurova [8] pointed out that the ice point varies in terms of the pavement types. Most ice prediction models are developed based on laboratory tests, and many significant factors are found to be related to ice generation. For example, Mohseni and Symons [9], and Diefenderfer et al. [10] both regressed the relationship between pavement temperature and various environmental conditions, such as illumination, air temperature, longitude, latitude, etc., but the impact of these factors is still unmeasurable. Thus, based on the existing models which were built by laboratory tests, the precise road surface condition is hard to be predicted.

Furthermore, several sensing technologies were developed for winter road surface condition monitoring. DSC-111 and DST-111 sensors are two remote optical sensors developed by Vaisala company [11–13]. DSC-111 can provide the road surface state (dry, moist, wet, icy, snowy/frosty or slushy) based on the backscattered signals of infrared light and can measure the friction level of the road surface and DST-111 can present the pavement surface temperature, air temperature and relative humidity by long wave infrared radiations detection [14]. Previous studies demonstrated DSC-111 can provide accurate surface state measurement, but the friction detection of DST-111 is not precise [12]. Road Condition Monitor (RCM) 411 is an optical instrument equipped with a transmitter to send a probe light pulse and a detector to measure the backscattered light, which can be easily installed to a passenger vehicle [15]. Existing researches conducted experiments to demonstrate that the RCM-411 is accurate in temperature, water thickness and road surface status detection [15–17]. For friction detection, even when the detected friction value does not always accurately match the actual friction, it still can be adjusted to the actual friction value based on calibration methods [15]. Such sensing technologies have already been employed for real-time road monitoring implementations, e.g. Road Weather Information Station (RWIS) in the US, etc. [16, 18–21]. However, each sensing technology has its own disadvantages, e.g. fixed sensor can only cover a fixed area, and using mobile sensors is time and energy consuming. Therefore, how to utilize the data collected by such sensing technologies for expanding ability and predicting the road surface condition would be valuable for improving the effectiveness and efficiency of the whole system.

By utilizing the data collected by existing sensing technologies, several researchers have developed data-driven prediction models for road surface condition related parameters forecasting. Liu developed a road surface temperature prediction model based on gradient extreme learning machine boosting algorithm [22]. Solol developed a road surface temperature prediction model based on energy balance and heat conduction models [23]. In addition, some researchers developed road surface condition recognition algorithms based on computer vision technologies [24–27]. However, previous studies have several disadvantages in terms of prediction effectiveness. For example, those methodologies can only regress the current road surface condition based on current environmental measurements, e.g. air temperature, etc. They are not able to predict road surface condition in the future. Moreover, researches in the past demonstrated the existence of the time-series features of road surface condition [28]. However, only a few studies shed light on the time-series prediction model development. Thus, a prediction method which considers the time-series features of road surface condition is needed based on the above analysis.

Long-short term memory neural network (LSTM) is a kind of computational intelligence approach for dealing with time-series data [29]. Previously, several studies demonstrated LSTM is more accurate in short-term prediction problems, e.g. traffic flow prediction, patient visitation frequency prediction, than other approaches, like random forest (RF) and support vector regression (SVR) due to the ability of handle both long-term and short-term dependencies. Based on the above considerations, the primary objective of this study is to develop a road surface friction prediction model based on the LSTM NN model using historical data. The RCM-411 friction sensing data were selected as the historical data set due to the accuracy. To evaluate the predictive effectiveness of the proposed method, several other prediction models were employed for the comparison purpose. Besides the overall prediction performance evaluation, the influence of the number of time-lags, the influence of the time interval between each time-step and the influence of adding additional features were also evaluated. Findings of this study can help to mitigate the impact of road surface condition on road traffic safety and mobility, especially in winter seasons.

## 2. Data

### 2.1. Testing Field

The data used in this study were collected by on-vehicle RCM 411 sensor on European route E75 from Sodankylä to Kemi in Finland. The total length of the road is 186 miles. In winter season, from October to next April, the air temperature is historically relatively low in this area. It could be minus 40 Celsius, and average minimum air temperature is about minus 15 Celsius. The average maximum air temperature is still under the ice point of water for the most of time. In the other seasons, the air temperature is not as high as in normal areas. The historical average maximum air temperature in July is about 20 Celsius. July is the month with the highest temperature in this area. Therefore, the study field has problems caused by cold weather, like icing and snow happened most frequently. There was a detection vehicle equipped with an RCM 411 sensor kept collecting the road surface friction data since February 2017. In addition, various road surface condition related parameters are sensed as well, e.g. calculated road surface status, water thickness, air temperature, etc., the detailed information would be introduced in the next section [30]. Figure 1 shows the distribution of calculated road surface status of two selected days. One is in winter season and one is in summer season. Basically, there are five calculated road surface status, including dry, moist, slush, ice, and snow or hoar frost. In the figure, calculated road surface status was variated along the road for both tow selected days. The most part of the road was covered by snow or hoar frost in winter season, while the most part of the road was dried in summer season.

### 2.2. Data Description

The historical data collected by the RCM 411 sensor in the introduced testing field was analysed in this study. The dataset covers 446 days from February 17th, 2017 to May 9th, 2018. During the time period, the vehicle equipped with the RCM 411 sensor drove through the testing field at least once per day, so that every point on the road has at least one friction record every single day. In this study, it is assumed that no spatial correlation of the road surface friction exists for adjacent road segments. Thus, the testing field was separated into road segments based on the calculated road surface status of the RCM 411 sensor. The road segmentation methodology will be introduced in section 3.1.

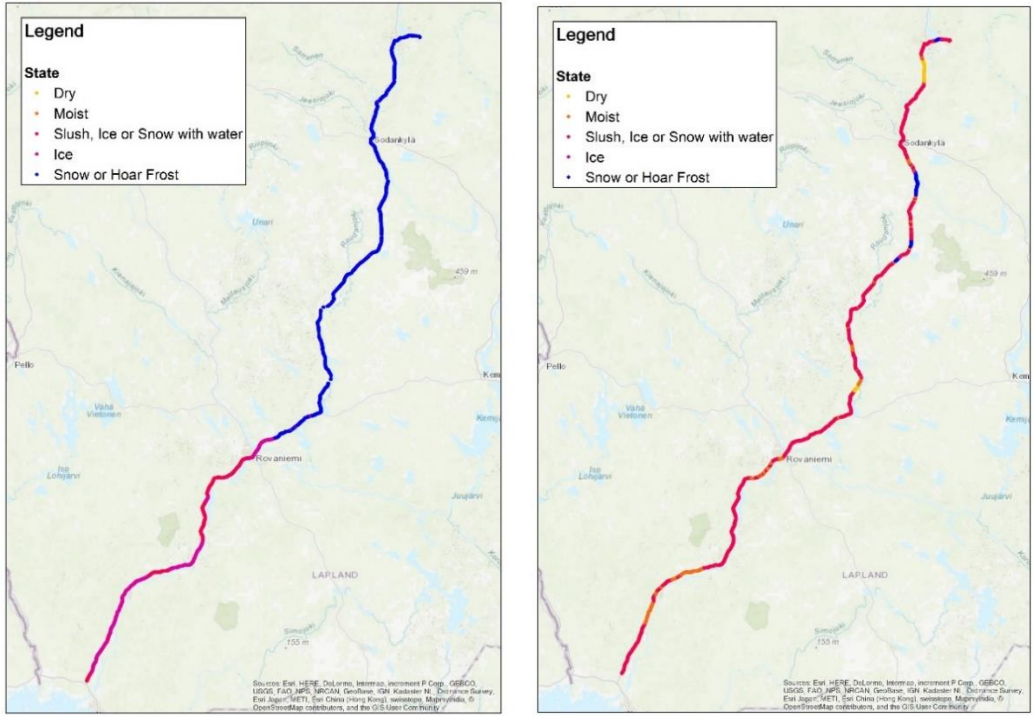

***Fig.1.*** *Road Surface Status Distribution: February 14th, 2018 (Left), and August 21st, 2017 (Right)*

As mentioned in section 1, the RCM 411 sensor can provide accurately calculated road surface status, air temperature data, road surface friction data, and road surface water thickness data. In most cases, road surface status is defined or measured based on the friction information, since the friction is directly related to traffic safety. In addition, the road surface friction coefficient is the most important indicator to characterize its anti-sliding performance which is an important indicator of the road safety quality. A stable road surface friction coefficient can provide a safety reserve for driving, thus reducing the possibility of traffic accidents occurrence. Therefore, the road surface friction is selected as the representative of the road surface condition in this study. Historical road surface friction data were used as the input of the proposed prediction model for predicting the road surface friction value in the future time period. Other historical data, like road surface water thickness, air temperature, and road surface temperature can also be used as the input of the prediction model for testing if they can improve the accuracy of the proposed prediction model.

## 3. Methodology

### 3.1. Road Segmentation

Road segmentation is the prerequisite for the prediction of road surface condition. Generally, the adjacent road segments share similar properties, but the distant sections are inclined to be different from each other. As mentioned in section 2.1, the whole distance of the study site is about 186 miles, and the calculated road surface status varies along with the road from dry to snow. The main objective of road segmentation is to guarantee that only one status exists within each road section during the testing time period. In order to make the prediction model comparable, the length of each segment should be the same. In this paper, we proposed a spatial clustering method based on the K-means clustering algorithm. K-means clustering partitions $n$ observations into $k$ clusters in which each observation belongs to the cluster with the nearest calculated road surface status. We used the spatial distance calculated by Haversine formula to describe the distance function instead of Euclidean distance, as shown in Eq. (1). In practice, we need to optimize the best $K$ number to improve the accuracy and reliability of the prediction model.

$$d = 2r\,arcsin\left(\sqrt{sin^2\left(\frac{\varphi_2 - \varphi_1}{2}\right) + cos(\varphi_1)\,cos(\varphi_2)\,sin^2\left(\frac{\lambda_2 - \lambda_1}{2}\right)}\right) \quad (1)$$

where $d$ represents the spatial distance, $r$ is radius of the earth, $\varphi_1$, $\varphi_2$ are the latitude of two points in radians, and $\lambda_1$, $\lambda_2$ are the longitude of two points, in radians.

### 3.2. Road Surface Friction Prediction using LSTM NN

A Long-short Term Memory (LSTM) neural network is proposed to predict the short-term road surface friction due to its ability to handle both long-term and short-term dependencies [31, 32]. LSTM shares similar architecture with traditional Recurrent Neural Networks (RNN), which are composed of one input layer, one hidden layer, and one output layer. The main modification of LSTM compared to RNN architecture is the structure of the hidden layer[33], which is shown in Figure 2.

Typically, at each time iteration $t$, the LSTM cell has the input layer, $X_t$, the output layer, $h_t$ and the hidden layer which is called LSTM cell. By adding a cell state component, the LSTM cell is capable of handling long-term dependencies of sequence data. The previous output cell state, $C_{t-1}$ and current input cell state, $\tilde{C}$, both influence the current output cell state, $C_t$. Three gates control the information to flow into and out of the cell state which are the forget gate, the input gate, and the output gate, denoted as $f_t$, $i_t$, and $o_t$, respectively. The forget gate controls how much information from previous cell state should be forgotten by the current cell state. The input gate handles how much information from the current input layer flows into the current cell state. The output gate controls how much information from the current cell state would be conveyed into the current output layer. They can be calculated by the following equations,

$$f_t = \sigma_g(W_f X_t + U_f h_{t-1} + b_f) \quad (2)$$

$$i_t = \sigma_g(W_i X_t + U_i h_{t-1} + b_i) \quad (3)$$

$$o_t = \sigma_g(W_o X_t + U_o h_{t-1} + b_o) \quad (4)$$

$$\tilde{C}_t = tanh(W_C X_t + U_C h_{t-1} + b_C) \quad (5)$$

where $W_f$, $W_i$, $W_o$, and $W_C$ are the weight matrices for mapping current input layer into three gates and current input cell state. $U_f$, $U_i$, $U_o$, and $U_C$ are the weight matrices for mapping the previous output layer into three gates and current input cell state. $b_f$, $b_i$, $b_o$, and $b_C$ are bias vectors for gate and input cell state calculation. $\sigma_g$ is the gate activation function which is normally a sigmoid function. $tanh$ is the hyperbolic tangent function which is the activation function for current input cell state. Then, the current output cell state and output layer can be calculated by the following equations. Finally, the output of the LSTM prediction model in this study should be the road surface friction in the next time iteration.

$$C_t = f_t * C_{t-1} + i_t * \tilde{C}_t \quad (6)$$

$$h_t = o_t * \tanh(C_t) \quad (7)$$

Since it is assumed no spatial correlation between road segments, the spatial dimension of the input data is set as $P = 1$. The unit of time-step for road surface friction detection is set as 1-day, then, the dataset has 446 time-steps for each road segment. Suppose the number of the time-lag is set as $T = t$ with $L = l$ days between each time-lag, which means the model used the data in previous $t$ consecutive time-steps to predict the road surface friction in the following 1 time-step. Then the dataset is separated into samples with $t$ time-lags and the sample size is $N = 446 - t$. Thus, each sample of the input data, $X_t$, is a 2-dimensional vector with the dimension of $[T, P] = [t, 1]$, and each sample of the output data is a 1-dimensional vector with 1 component. The input of the model for each road segment is a 3-dimensional vector which dimension is $[N, T, P] = [446 - t, t, 1]$. Before feeding into the model, all samples are randomly divided into three datasets for training, validating, and testing with the ratio 7:2:1.

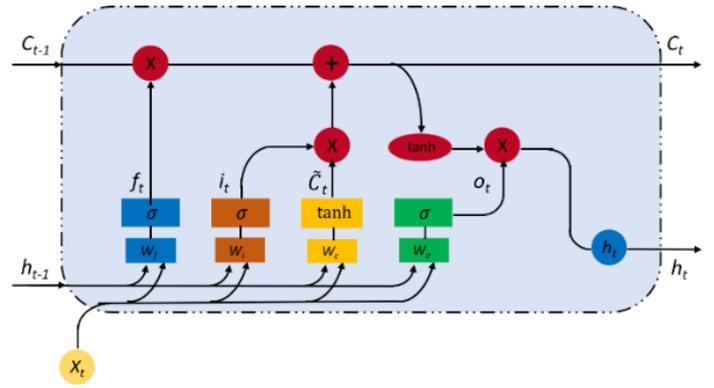

***Fig. 2.*** *Model Architecture of LSTM (Red circle are arithmetic operators and the rectangles in different colours are the gates in LSTM)*

### 3.3. Predictive Performance Evaluation

The performance of a LSTM NN in road surface friction prediction is compared to that of many classical baseline models for short-term prediction. Typically, ARIMA, Support Vector Regression (SVR), Random forest (RF), Kalman filter, tree-based model and feed-forward NN were used for addressing short-term prediction problems [34–36], e.g. traffic speed and travel time prediction [37–39]. However, several time-series prediction models were demonstrated that the predictive performance is not as accurate as others, e.g. ARIMA and Kalman filter. Therefore, based on previous research results, SVR, RF and feed-forward NN were selected for comparing the performance of road surface friction prediction with the proposed LSTM NN model in this study. Among these models, feed-forward NN, which is also called Multilayer Perceptron, is popular for precise performance in short-term prediction [40]. RF and SVR are also well-known models for efficient predictive performance [34, 35]. For the parameters of model development, the Radial Basis Function (RBF) kernel is deployed in the SVR model. 10 trees were built, and there was no pre-determined limitation for the maximum depth of the trees for the RF model. The feed-forward NN was composed of 2 hidden layers with 100 nodes in each layer.

Mean Absolute Error (MAE), Mean Square Error (MSE) and Mean Absolute Percentage Error (MAPE) are used as the measurements of predictive performance. The following equations present the measurement formulation.

$$MAE = \frac{\sum_{i=1}^{N}|Y_i - \hat{Y}_i|}{N} \quad (8)$$

$$MSE = \frac{\sum_{i=1}^{N}(Y_i - \hat{Y}_i)^2}{N} \quad (9)$$

$$MAPE = \frac{100\%}{N}\sum_{i=1}^{N}\left|\frac{Y_i - \hat{Y}_i}{Y_i}\right| \quad (10)$$

where $N$ is the total number of samples in testing date set, $Y_i$ is the ground truth of the road surface friction which is detected by RCM 411 sensor in this study, and $\hat{Y}$ is the predicted road surface friction of the proposed prediction model. Typically, the MAE is used to measure the absolute error associated with a prediction, the MAPE presents a measure of the percentage of average misprediction of the model and the MSE measures the relative error for a prediction. The prediction model with the smaller values of MAE, MSE, and MAPE performs better.

## 4. Numerical Results

### 4.1. Road Segmentation

As shown in Eq. (1), a large $K$ value lowers the size of each cluster, leading to insufficient data to train the LSTM model. In contrast, a small $K$ value may result in a mixture of various types of status. Therefore, we proposed a heuristic method by slightly increasing the $K$ and evaluating the degree of mixture of each cluster. However, it is practically impossible to ensure only one status existed in each segment. Even for the 100-meter road segment, there could be a 1-meter section with another status, which is physically normal. Thus, the proportion between the outlier status and the normal status of each segment is used as an indicator (mixture rate), and the $K$ is derived to guarantee the mixture rate below 15%. Finally, the $K$ is calculated as 1487. Figure 3 visualizes the calculated road surface status of the randomly selected road segments on two days. Seen from the figures, even the calculated road surface statuses are different, only one status exists within the selected road segments. After segmenting the road based on the proposed criteria, the average road surface friction could be used to represent the value of each road segment. Otherwise, the value of the road surface friction within each road segments could vary a lot.

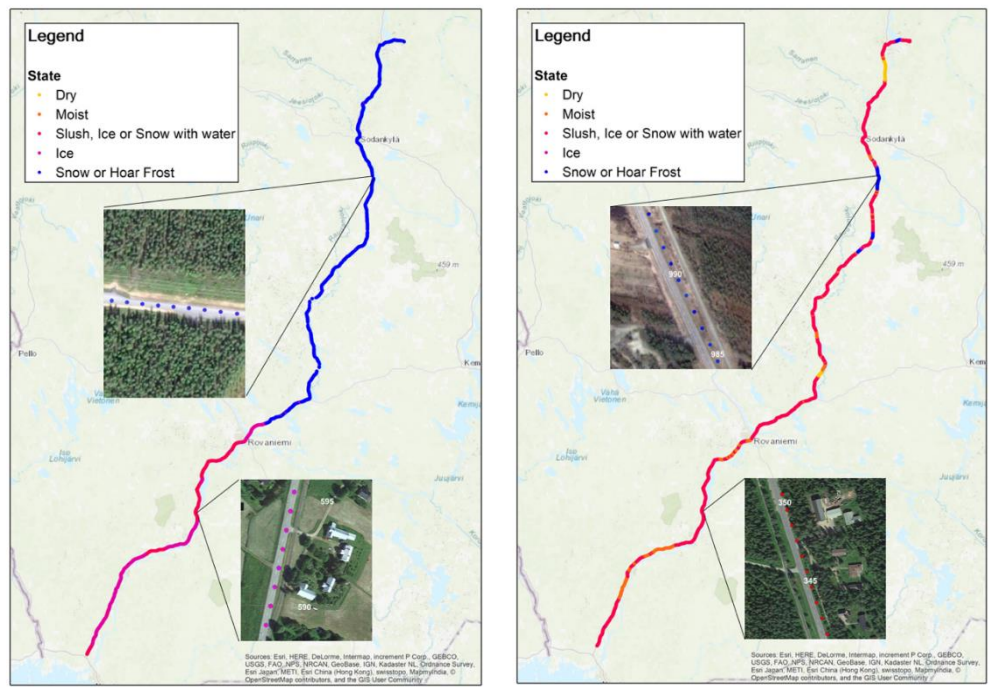

***Fig.3.*** *The Road Segmentation Results*

### 4.2. Predictive Performance Evaluation

The proposed LSTM NN model and other baseline models were trained based on the same training data set for each road segment separately, and the predictive performance for each model was calculated based on predicted value and ground truth value. In this step, only road surface friction in the previous time period was used as the model input. The final predictive performance measurements were the averaged value of all road segments. Table 1 shows the prediction performance comparison of the LSTM with other baseline models. Among other algorithms, RF performed much better than SVR and feed-forward NN with MAE of 0.166, MSE of 0.0132 and MAPE of 16.6%, which makes sense due to the majority votes mechanism of RF model. The feed-forward NN had the worst predictive performance, which is caused by the sparsity of the data. The proposed LSTM model outperformed all models with MAE of 0.0778, MSE of 0.0112 and MAPE of 15.16%, which indicates the best performance in predicting road surface friction by only consider the road surface friction in the previous time period.

To further examine the predictive performance of the proposed LSTM model in a more intuitive way, the comparison of the predicted values of the LSTM on a randomly selected day for all road segments were compared to the ground truth value, which is presented in Figure 4. For most of road segments, the predicted values were very close to the observed data. Only a few of the road segments had clear errors in road surface condition prediction. Overall, the LSTM effectively predicted road surface friction based on historical road surface friction data for all road segments.

### 4.3. Evaluating the Influence of Number of Time-Lags on Predicting Accuracy

The number of time-lags is the temporal dimension of the input data, which might influence the prediction performance of the proposed LSTM model. Intuitively, the more time-lags will convey temporal features in a longer time period, and the LSTM will learn more features in previous time periods. In order to explore the influence of the number of time-lags, the LSTM was trained by the data sets with a different number of time-lags, from 1 to 10 separately, for all road segments. All data samples had the same time interval (1-day) between time-lags. Table 2 shows the average predictive performance of the proposed LSTM models which were trained by the data sets with a different number of time-lags.

It is noticed that all three measurements (MAE, MSE, and MAPE) gradually dropped from the number of time-lags equalling 1 to 7. The LSTM model performed with the most precise prediction when the number of time lags equals to 7. Once the number of time lags was greater than 7, the prediction performance became worse with a little fluctuation. The potential reason might lie on the excessive time lags made the LSTM too complex, which caused some overfitting issues with the LSTM. Thus, the prediction effectiveness was influenced by the unnecessary complexity of the LSTM.

### 4.4. Evaluating the Accuracy of the Prediction after Different Days

The time interval between time-lags indicates how often a historical data point will be input into the proposed LSTM model. In this study, the frequency of road surface friction detection is once per day, then the minimum time interval between time-lags is 1-day. If the time interval between time-lags is set as 1-day, then the output would be the road surface friction after 1-day. Thus, if the road surface friction after $l$ days is predicted, the time interval between each time-lags of the input data should be set as $l$. By varying the time interval, the prediction time can be adjusted. Then, the model is not only dedicated to predicting the road surface friction after a fixed number of days. In order to demonstrate the road surface friction prediction accuracy after different days, the proposed LSTM was trained separately by the data sets with a different time interval between time-lags from 1 to 10 for all road segments. Table 3 shows the average predictive performance of the LSTM models.

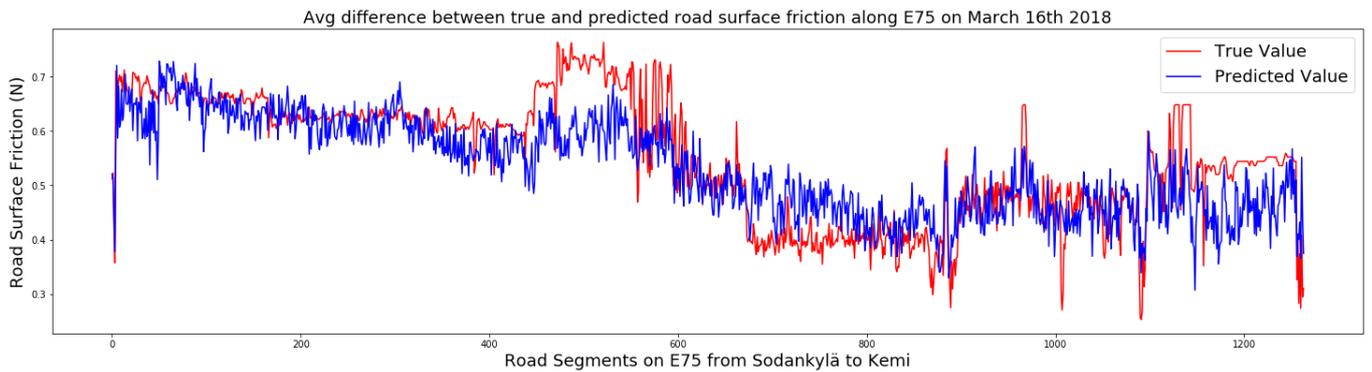

***Fig.4.*** *Predictive Performance Comparison of the LSTM with Observed Data*

**Table 1** Predictive Performance Comparison of the LSTM with Other Models

| Models | MAE (N) | MSE | MAPE (%) |
|---|---|---|---|
| Feed-forward NN | 0.1660 | 0.0132 | 26.86 |
| SVR | 0.2142 | 0.0174 | 21.42 |
| RF | 0.1660 | 0.0132 | 16.60 |
| LSTM NN | 0.0778 | 0.0112 | 15.16 |

**Table 2** Predictive Performance of the LSTM with Different Number of Time Lags

| Time Lages | 1 | 2 | 3 | 4 | 5 | 6 | 7 | 8 | 9 | 10 |
|---|---|---|---|---|---|---|---|---|---|---|
| MAE (N) | 0.0862 | 0.0836 | 0.0787 | 0.0812 | 0.0799 | 0.0800 | 0.0778 | 0.0832 | 0.0797 | 0.0824 |
| MSE | 0.0135 | 0.0133 | 0.0117 | 0.0126 | 0.0119 | 0.0121 | 0.0112 | 0.0128 | 0.0117 | 0.0126 |
| MAPE (%) | 17.62 | 17.82 | 16.23 | 16.97 | 16.14 | 16.57 | 15.16 | 16.81 | 15.58 | 16.65 |

**Table 3** Predictive Performance of the LSTM with Different Interval between Time Lags

| Time Interval (Days) | 1 | 2 | 3 | 4 | 5 | 6 | 7 | 8 | 9 | 10 |
|---|---|---|---|---|---|---|---|---|---|---|
| MAE (N) | 0.0790 | 0.0873 | 0.0903 | 0.0948 | 0.1043 | 0.1096 | 0.1000 | 0.1086 | 0.1085 | 0.1190 |
| MSE | 0.0127 | 0.0146 | 0.0152 | 0.0166 | 0.0195 | 0.0229 | 0.0189 | 0.0222 | 0.0220 | 0.0232 |
| MAPE (%) | 15.24 | 18.01 | 17.61 | 19.99 | 21.06 | 21.75 | 19.86 | 22.63 | 21.33 | 22.39 |

Obviously, as the time interval between time lags became larger, the predictive performance of the LSTM got worse for all three performance measurements. It suggested the accuracy of road surface friction prediction would be decreased when the predicting interval getting larger. It is noticed that, as the predicting interval became larger, the prediction accuracy did not drop too much to make the prediction accuracy unacceptable. The road surface friction prediction of 5 days later still got about 20% MAPE and relatively low MSE and MAE. Even when the time interval between time lags equals 10 days, the proposed LSTM model still got 22.39% MAPE. Figure 5 shows the boxplots of the predictive performance of the proposed LSTM models trained by the data with different days between time lags. It is explicit that, while the time interval between time-lags became large, the variance of predictive performance was getting larger for all three predictive performance measurements. The 25th percentiles of three measurements were stable as the predicting time interval getting larger. The 75th percentile increased while the predicting time interval is set from 1day to 10 days. In summary, the proposed LSTM model is accurate for predicting short-term road surface friction. When the predicting time interval becomes larger, the prediction accuracy decreases, which is consistent with the previous research results that the road surface weather condition has short-term time-series features but long-term features[41].

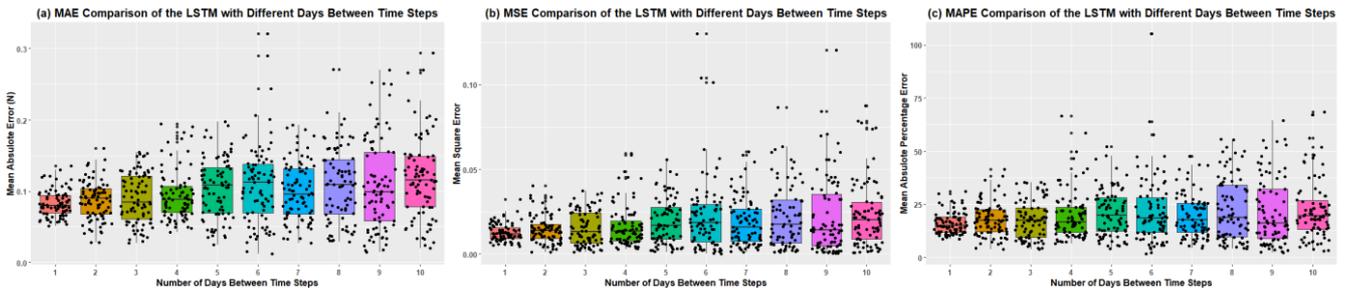

***Fig.5.*** *Boxplots of the Predictive Performance of The LSTM with Different Days Between Time Lags*
*(a)* MAE Comparison of the LSTMs with Different Days Between Time Steps, *(b)* MSE Comparison of the LSTMs with Different Days Between Time Steps, *(c)* MAPE Comparison of the LSTMs with Different Days Between Time Steps

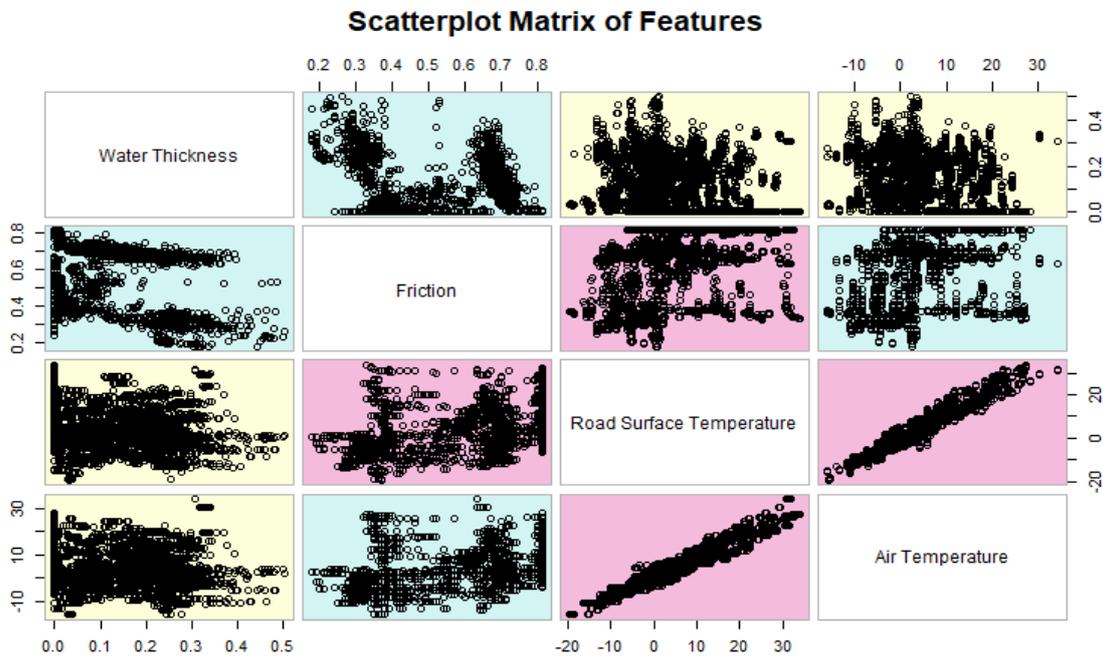

***Fig.6.*** *Scatter Plots Matrix of Features*

### 4.5. Evaluating the Influence of Other Related Features on Predicting Accuracy

The above LSTM prediction models were trained only by friction value in the past time periods. Theoretically, the road surface friction would be mainly determined by road surface water thickness, road surface temperature, and air temperature. Thus, it is meaningful to add more variables as the additional input of the LSTM model to explore the influence of those features to the prediction accuracy.

Figure 6 shows the scatterplot matrix of road surface water thickness, road surface friction, road surface temperature and air temperature collected by RCM 411 sensor to display the correlation among these features. The road surface temperature and air temperature have a heavily strong correlation that all the dots centralized to the diagonal line. However, road surface friction seems does not have a clear correlation with two temperature related measurements. The dots spread in the plots without specific patterns. In addition, the scatter plot of road surface water thickness and road surface friction presents a U-shaped pattern. The road surface water thickness reached a large value when the road surface friction value is relatively large or small. Based on the above consideration, two additional experiments were conducted for investigating the influence of these features on predicting accuracy. The LSTM models were trained by adding road surface water thickness and adding road surface water thickness and temperature. 7 time-lags and the 1-day time interval between time-lags were selected for the model training. The prediction performance was compared with the prediction performance of the LSTM model trained only by road surface friction. The comparison result is shown in Table 4.

As shown in Table 4, the predictive performance of the LSTM was improved by adding road surface water thickness as additional input of the model. All three predictive performance measurements achieved lower value. It is demonstrated the statement in the previous study that road weather condition correlates with the rainfall in the past time period [42]. However, when road surface water thickness and road surface temperature were added as an additional input of the prediction model, the predictive performance became worse than only taking road surface friction as the input. All three performance measurements increase a lot. By considering the weak correlation between road surface friction and temperature related measurements, the potential reason could be that the additional temperature related features made the LSTM too complicated and brought lots of useless information to the LSTM model. The weights effectiveness of useful feature could be influenced by the excessively complex model structure, thus reducing the model accuracy. In the previous study, the same situation was found for short-term traffic speed prediction [38]. In that study, the accuracy of traffic speed prediction of the proposed LSTM was not improved by adding traffic volume and traffic occupancy as additional features. In summary, the accuracy of the proposed LSTM prediction model was improved by combined road surface water thickness in the past time period as additional input for predicting the road surface friction after 1 day, but the accuracy was weakened by adding road surface water thickness and temperature simultaneously as the additional input due to the excessively complicated model structure.

**Table 4** Predictive Performance Comparison of the LSTM with Different Features

| Data Input of Prediction Model | MAE (N) | MSE | MAPE (%) |
| --- | --- | --- | --- |
| Friction | 0.0778 | 0.0112 | 15.16 |
| Friction, Water Thickness | 0.0742 | 0.0102 | 14.58 |
| Friction, Water Thickness, Road surface Temperature | 0.0948 | 0.0159 | 21.97 |

## 5. Conclusion and Future Work

This study employed LSTM NN to develop a road surface friction prediction model based on historical friction data. The road surface friction data on European route E75 from Sodankylä to Kemi collected by RCM 411 sensor was used as model input, which covered 446 days and over 186 miles in total. The road was segmented into 1,487 road segments based on the calculated road surface status. The experiments were conducted for each road segments independently, and the predictive performance of the proposed LSTM was calculated by averaging the predictive performance measurements of all road segments prediction results. For demonstrating the effectiveness of the proposed model, SVR, RF and Feed-forward NN were selected as the baseline models for comparing the predictive performance with the proposed model. Furthermore, the impact of the number of time-lags on predictive accuracy, the influence of time interval between time steps on predictive accuracy were also tested. In addition, the experiments also analyzed the impact of adding road surface water thickness, road surface temperature and air temperature on predictive accuracy.

Based on the analysis results, the proposed LSTM road surface friction prediction model outperformed all other baseline models in terms of the lowest value of MAE, MSE, and MAPE. The proposed LSTM prediction model got 0.0778 in MAE, 0.0112 in MSE and 15.16% in MAPE. The number of time-lags and the predictive time interval had an influence on the predictive performance of the proposed model. The LSTM prediction model achieved the most accurate prediction with 7 time-lags, and the prediction accuracy dropped when the predictive time interval was getting larger. Road surface water thickness and road surface temperature were added to the proposed prediction model as additional model input. Road surface water thickness improved the predictive accuracy, but road surface temperature did not. The findings of this study can be used to support the road maintenance plan and decision making, thus mitigating the impact of inclined inclement road surface condition on traffic safety and mobility. In the future, an improved LSTM prediction model should be developed for freeing the requirement of a fixed time interval between each time-lags of one data sample. Thus, the energy and time cost for data collection can be saved, which makes the prediction model more convenient and valuable from an implementation perspective.

## 6. Acknowledgments

This research was supported by the multi-institutional project (Exploring Weather-related Connected Vehicle Applications for Improved Winter Travel in Pacific Northwest) of Pacific Northwest Transportation Consortium (PacTrans) USDOT University Transportation Center for Federal Region 10.## 7. References

[1] Pisano, P.: 'and road weather'2017, (August), pp. 144–145.

[2] Ye, Z., Strong, C.K., Shi, X., Conger, S.M., Huft, D.L.: 'Benefit-cost analysis of maintenance decision support system'Transp. Res. Rec., 2009, (2107), pp. 95–103.

[3] Chen, S., Saeed, T.U., Labi, S.: 'Impact of road-surface condition on rural highway safety: A multivariate random parameters negative binomial approach'Anal. Methods Accid. Res., 2017, 16, pp. 75–89.

[4] Shi, X.: 'Winter Road Maintenance: Best Practices, Emerging Challenges, and Research Needs'Transp. Res. Rec., 2011, 2, (4), pp. 1–5.

[5] Rita, U.: 'PacTrans Region 10 University Transportation Center Exploring Weather-related Connected Vehicle Applications for Improved Winter Travel in Pacific Northwest'2018.

[6] Wallman, C.-G., Åström, H.: 'Friction measurement methods and the correlation between road friction and traffic safety: A literature review' (Statens väg-och transportforskningsinstitut, 2001)

[7] Shao, J., Lister, P.J., Shao, J., Lister, P.J.: 'An Automated Nowcasting Model of Road Surface Temperature and State for Winter Road Maintenance' (1996), 35, pp. 1352–1361

[8] Samodurova, T. V: 'Estimation of significance the parameters , influencing on road ice formation ( the results of computing experiment )'no date.

[9] Contact, F., Symons, M.: 'TECH BRIEF LTPP Data Analysis : Improved Low Pavement Temperature Prediction'1995, (703).

[10] Diefenderfer, B.K., Asce, A.M., Al-qadi, I.L., et al.: 'Model to Predict Pavement Temperature Profile : Development and Validation'2006, 132, (2), pp. 162–167.

[11] Ye, Z., Shi, X., Strong, C.K., Larson, R.E.: 'Vehicle-based sensor technologies for winter highway operations'IET Intell. Transp. Syst., 2012, 6, (3), p. 336.

[12] F, Feng., Liping, Fu., 'Evaluation of Two New Vaisala Sensors for Road Surface Conditions Monitoring Final Report'2008, (August).

[13] Ewan, L., Al-Kaisy, A., Veneziano, D.: 'Remote Sensing of Weather and Road Surface Conditions'Transp. Res. Rec. J. Transp. Res. Board, 2013, 2329, (2329), pp. 8–16.

[14] Pilli-sihvola, Y., Toivonen, K., Haavasoja, T., Haavisto, V., Nylander, P.: 'March , Turin , ITALY New Approach to Road Weather : Measuring Slipperiness March , Turin , ITALY'2006, pp. 13–18.

[15] Haavasoja, T., Nylander, J., Nylander, P.: 'Experiences of Mobile Road Condition Monitoring'Proc. 16th Int. Road Weather Conf., 2012, (May), pp. 23–25.

[16] Maenpaa, K., Sukuvaara, T., Ylitalo, R., Nurmi, P., Atlaskin, E.: 'Road weather station acting as a wireless service hotspot for vehicles'Proc. - 2013 IEEE 9th Int. Conf. Intell. Comput. Commun. Process. ICCP 2013, 2013, pp. 159–162.

[17] Fay, L., Akin, M., Muthumani, A.: 'QUANTIFYING SALT CONCENTRATION ON PAVEMENT – PHASE II'2018.

[18] Karsisto, V., Nurmi, P.: 'Using car observations in road weather forecasting 18th international Road Weather Conference'2016, (April).

[19] Singh, G., Bansal, D., Sofat, S., Aggarwal, N.: 'Smart patrolling: An efficient road surface monitoring using smartphone sensors and crowdsourcing'Pervasive Mob. Comput., 2017, 40, pp. 71–88.

[20] Saarikivi, P.: 'Development of mobile optical remote road condition monitoring in Finland'2012, (May), pp. 23–25.

[21] Boselly, E., Thornes, E., Ulburg, C., Ulburg, C.: 'Road weather information systems volume 1: Research report'Strateg. Highw. Res., 1993, pp. 90–93.

[22] Liu, B., Yan, S., You, H., et al.: 'Road surface temperature prediction based on gradient extreme learning machine boosting'Comput. Ind., 2018, 99, (March), pp. 294–302.

[23] Sokol, Z., Bližňák, V., Sedlák, P., Zacharov, P., Pešice, P., Škuthan, M.: 'Ensemble forecasts of road surface temperatures'Atmos. Res., 2017, 187, pp. 33–41.

[24] Cnn, P.: 'Winter Road Surface Condition Recognition Using a Pre-trained Deep Convolutional Neural Network'no date, p. 16.

[25] Sukuvaara, T., Nurmi, P.: 'ID : 001 Connected vehicle safety network and road weather forecasting – The WiSafeCar project'2012, (May), pp. 23–25.

[26] Linton, M.A., Fu, L.: 'Connected Vehicle Solution for Winter Road Surface Condition Monitoring'Transp. Res. Rec. J. Transp. Res. Board, 2016, 2551, pp. 62–72.

[27] Jonsson, P., Edblad, J., Thörnberg, B.: 'Developing a cost effective multi pixel NIR camera for road surface status classification in 2D'2014, pp. 1–6.

[28] Kangas, M., Heikinheimo, M., Hippi, M.: 'RoadSurf: a modelling system for predicting road weather and road surface conditions'Meteorol. Appl., 2015, 22, (3), pp. 544–553.

[29] Hochreiter, S., Urgen Schmidhuber, J.: 'Long Short-Term Memory'Neural Comput., 1997, 9, (8), pp. 1735–1780.

[30] 'User ' s Guide Road Condition Monitor RCM411'no date.

[31] Bengio, Y., Simard, P., Frasconi, P.: 'Learning Long-Term Dependencies with Gradient Descent is Difficult'IEEE Trans. Neural Networks, 1994.

[32] Sundermeyer, M., Schl, R., Ney, H.: 'LSTM Neural Networks for Language Modeling'Proc. Interspeech, 2012, pp. 194–197.

[33] Gers, F. a, Cummins, F.: '1 Introduction 2 Standard LSTM'1999, pp. 1–19.

[34] Wu, C., Wei, C., Su, D., Chang, M., Ho, J.: 'Travel time prediction with support vector regression'Proc. 2003

IEEE Int. Conf. Intell. Transp. Syst., 2004, 2, pp. 1438–1442.

[35] Yuan-yuan Chen, Lv, Y., Li, Z., Wang, F.-Y.: 'Long short-term memory model for traffic congestion prediction with online open data'2016 IEEE 19th Int. Conf. Intell. Transp. Syst., 2016, pp. 132–137.

[36] Guo, J., Huang, W., Williams, B.M.: 'Adaptive Kalman filter approach for stochastic short-term traffic flow rate prediction and uncertainty quantification'Transp. Res. Part C Emerg. Technol., 2014, 43, pp. 50–64.

[37] Ma, X., Tao, Z., Wang, Y., Yu, H., Wang, Y.: 'Long short-term memory neural network for traffic speed prediction using remote microwave sensor data'Transp. Res. Part C Emerg. Technol., 2015, 54, pp. 187–197.

[38] Cui, Z., Ke, R., Wang, Y.: 'Deep Bidirectional and Unidirectional LSTM Recurrent Neural Network for Network-wide Traffic Speed Prediction'2018, pp. 1–12.

[39] Cui, Z., Henrickson, K., Ke, R., Wang, Y.: 'High-Order Graph Convolutional Recurrent Neural Network: A Deep Learning Framework for Network-Scale Traffic Learning and Forecasting'2018, pp. 1–9.

[40] Lv, Y., Duan, Y., Kang, W., Li, Z., Wang, F.Y.: 'Traffic Flow Prediction With Big Data: A Deep Learning Approach'IEEE Trans. Intell. Transp. Syst., 2014, 16, (2), pp. 865–873.

[41] Brijs, T., Karlis, D., Wets, G.: 'Studying the effect of weather conditions on daily crash counts using a discrete time-series model'Accid. Anal. Prev., 2008, 40, (3), pp. 1180–1190.

[42] Hambly, D., Andrey, J., Mills, B., Fletcher, C.: 'Projected implications of climate change for road safety in Greater Vancouver, Canada'Clim. Change, 2013, 116, (3–4), pp. 613–629.